# QBism and relational interpretation of quantum mechanics from the point of view of a contextual quantum realism (CQR)[1]

## François-Igor Pris (И. Е. Прись)

A realist interpretation of quantum mechanics is proposed – Contextual Quantum Realism (CQR) – according to which there exists a categorical distinction between the ideal (theory, observation instrument) and the real (quantum physical systems, properties), and, consequently, quantum ontology is context-sensitive. CQR is compared with QBism and Relational Quantum Mechanics (RQM), both of which also claim to offer realist interpretations. However, both approaches conflate the ideal and the real, thereby introducing an anti-realist dimension. RQM is an objectivist/physicalist interpretation that naturalizes epistemic concepts, whereas QBism is subjectivist/phenomenological. RQM and QBism share a common presupposition of (post-)Modern philosophy – namely, the reduction of reality to objectivity – which CQR explicitly rejects.

## 1 Introduction

In our articles and books, we've argued that quantum mechanics (QM), as a radically new (non-classical) physical theory (paradigm), requires an equally radical new (non-classical) philosophical paradigm for its understanding. However, since its inception, all its numerous interpretations, both anti-realistic and realistic (broadly speaking), have been confined within the paradigm of Modern philosophy and/or post-Modern philosophy, which we see as a reversal (or limiting case) of Modern philosophy. This is the root cause of the quantum paradoxes, for which solutions have been proposed for almost a century, but none have been satisfactory.

In our view, this new philosophical paradigm is contextual realism (CR), a framework with Wittgensteinian origins that emerged at the beginning of the 21st century and continues to be developed and applied through the work of J. Benoist and our own efforts. CR rejects the premises and core characteristics of (post-)Modern philosophy, particularly representationalism, subjectivism, internalism, transcendentalism, perspectivism, and phenomenology, which treats sensory experience as a (true or false) phenomenon [Benoist, 2017; Benoist, 2023; Benoist, 2024; Pris, 2020; Pris, 2022; Pris, 2023]. We've termed a new interpretation of QM from the perspective of CR as contextual quantum realism (CQR) [Pris, 2020; Pris, 2022; Pris, 2023]. We've asserted that CQR can resolve or therapeutically "dissolve" the paradoxes of quantum mechanics. In a sense, this was a return to the origins: the "Copenhagen interpretation" of N. Bohr, W. Heisenberg, M. Born, W. Pauli, J. von Neumann, and others (whose philosophical views on quantum mechanics generally differed), which was considered anti-realistic but is re-evaluated in a new light.

---

[1] This work was partially supported by the Belarusian Republican Foundation for Fundamental Research, project Г24МС-002: "Quantum-like Modeling of Social and Human Systems and Philosophy of Contextual Realism".

In our opinion, the interpretation of the philosophical views of the founding fathers of QM cannot be unambiguous. We proposed an interpretation of Heisenberg's position and his concept of a closed theory from the perspective of CR [Pris, 2023]. We also argued that Bohr's approach can be understood as both neo-Kantian in a broad sense and contextual and realistic from the perspective of CR [Pris, 2023]. The emergence of interpretations that were realist in the traditional metaphysical sense (many-worlds, collapse, with hidden variables), while useful in a heuristic and scientific sense and expressing a natural dissatisfaction with a purely instrumentalist approach and the "orthodox" Copenhagen interpretation, was philosophically a step backward compared to it. This explains the opinion of M. Bitbol, with whom, as follows from the above, we only partially agree, that even before the quantum "paradoxes" were explicitly formulated, they were "dissolved" by N. Bohr and W. Heisenberg, thanks to the replacement of "the formal concepts of intrinsic properties, and substrates of these properties, by the formal concept of relational 'observables' and entities. <…> The post-quantum mutation of the formal concepts of physics thus leads to the death of an old ontology coupled with an old epistemology, and simultaneously to the birth of a new ontology coupled with a new epistemology" [Bitbol, 2024, p. 409].

Thus, the emergence of neo-Copenhagen and Copenhagenish interpretations in the 21st century can also be seen as both a return to the origins and a step forward [Schmid, Ying, Leifer, 2025]. At the same time, unlike CQR, they are all – informational, pragmatist, QBism, relational, and others – again based on some premises of (post-)Modern philosophy. In particular, we've argued that A. Zeilinger's informational interpretation, based on the principle of information, is anti-realistic. We've proposed replacing this principle with the principle of contextuality [Pris, 2022, part 9, ch. 4; Pris, 2023]. Pragmatist interpretations fit within the position of normative pragmatism and take a step forward in understanding QM, but they also cannot be considered realistic (for example, R. Healey refers to the neo-Hegelian R. Brandom). Finally, as we've argued, C. Fuchs's QBism and C. Rovelli's relational quantum mechanics (RQM) are two extreme positions within the Modern philosophy, which, strictly speaking, cannot be considered genuinely realistic interpretations (although they claim to be) [Pris, 2023]. QBism is a subjectivist interpretation [Adlam, Rovelli, 2022, web; Fuchs, 2023, web; Rovelli, 2021a; Rovelli, 2021b; Schack, 2023, web]. There are also phenomenological and transcendental interpretations of this interpretation, which M. Bitbol, for example, sympathizes with. But as we have shown, they also cannot be considered satisfactory [Pris, 2023, ch. 5-6]. RQM, on the contrary, is an objectivist, naturalistic/physicalist interpretation in the traditional sense [Pris, 2023, ch. 2].

We've demonstrated the possibility of correcting and re-interpreting QBism and RQM within the framework of CQR. According to us, both conflate the categories of the ideal and the real (positions that allow such a conflation we classify as anti-realistic/idealistic), and violate the grammar of the concept of fact by introducing the concept of relative facts instead of acknowledging the contextuality of facts in the sense that CQR understands it (the context is not something external to the judgment, rule, norm, or concept that identifies a fact, or to the identified fact itself). In this paper, we provide additional arguments and insist on our critique of both QBism and RQM, as well as Bitbol's position.

In paragraph 2, we restate the main tenets of CQR; in paragraph 3, we offer an interpretation of QBism from the perspective of CQR; in paragraph 4, RQM is considered from the perspective of CQR. In paragraph 5, we conclude in favor of CQR. When examining QBism and RQM, we compare them with each other and with CQR, as well as, briefly, with Bitbol's viewpoint.

## 2 CQR

CQR interprets quantum theory as a Wittgensteinian rule (norm, conceptual scheme) rooted in experience (hereafter, W-rule). This W-rule is for measuring reality within the "language games" (LGs) of its applications. LGs are normative and contextual. What exists – the quantum ontology – is identified within these LGs and, therefore, is sensitive to context and not predetermined [Pris, 2020; Pris, 2022; Pris, 2023; Pris, 2024].

By rejecting representationalism, CQR dismisses the metaphysical scientific realism of the "external world" with predetermined objects and their relations, which are described (represented) by a theory acting as a mediator between them and the subject, providing access to them. (In a Kantian sense of representation, the theory only provides access to phenomena, not the things themselves).

By rejecting phenomenology, CQR rejects the treatment of the primary sensory experience as a phenomenon, an appearance (which either shows itself or implies something hidden, of which it is an appearance). It rejects, in other words, the epistemologization of the sensory, particularly perceptual (visual), experience. The sensory (in French: le sensible) is real [Benoist, 2024]. While phenomenology asserts that appearance is "reality" and there is no distance between phenomenon (appearance) and being, CQR argues that there is no distance between the non-conceptualized sensory experience and reality, and that appearance is secondary and has a normative structure: it presupposes a judgment (which is always contextual) and is governed by the norm of truth.

Real things are not reduced to objects – things that are known or knowable, fully or partially – and they are not something unknowable. The grammar of the concept of reality is as follows: "Reality is just *what it is*" [Benoist, 2014, p. 22]. That which is correctly identified by the W-rule, and therefore that about which one can be mistaken, is real. Benoist, citing J.L. Austin's research into the multidimensionality of the concept of the real, writes that this concept "can be applied to things of very different kinds, depending on the different ways in which they may not be themselves – which, however, implies, as one can notice, that each time it is a matter of things for which it makes sense to be themselves, which may be a good enough positive definition of the real" [Benoist, 2024, p. 241–242].

By rejecting transcendentalism, CQR dismisses (neo-)Kantian and, more broadly, subject-object correlationism of Modern philosophy, which asserts only the possibility of knowing what is already known—the relationship between subject and object. It also rejects perspectivism, which structurally limits knowledge to a correlation or a perspective. (Thus, neo-Kantian and perspectivist interpretations of quantum mechanics are rejected). Methodologically, CQR is a grammatical approach. "Grammatical analysis, as Benoist writes, far from being a re-enactment of the transcendental, should normally defuse both the ontological mythology of the "thing in itself" and the transcendental mythology of the 'I_think'" [Benoist, 2023, p. 133].

CQR also asserts a categorical (not a substantial) dualism of the ideal and the real. Quantum theory, as a W-rule (norm), is ideal, not part of reality. The quantum physical quantities whose values are identified by this rule are real. The result of a quantum measurement is not predictable or predetermined because it is contextual; it arises in its own individual context, which is generally incompatible with the contexts of other measurements. In other words, quantum ontology (and, in fact, according to CR, any ontology) is sensitive to context, and is therefore contingent, not necessary. CQR thus supports contingentism against necessitism [Williamson, 2015].

CQR asserts that what is measured is as it is, as it was before its measurement, and as it will be after its measurement (realism: the measurement does not produce its result). However, before its measurement, it made no sense to say that it had a definite value, that is, before the measurement it had no identity (against correlationism or objectivism: the measurement does not measure an already measured, known, observed, or (pre)determined result). Traditional realists and anti-realists share a common premise of (post-)Modern philosophy that reduces reality to objectivity. Anti-realists assume that since the result of a measurement is not predetermined, it is created. In contrast, realists assume that because the result is real and not created, it must be predetermined.

## 3 CQR vs. QBism

There's an important similarity between our CQR (Contextual Quantum Realism) and QBism (Quantum Bayesianism) [Fuchs, 2023, web; Schack, 2023, web; Pienaar, 2021, web]. QBism also asserts that quantum measurement is an action. But for QBism, this is an action that the quantum agent performs on the "external world"; it is an interaction with the external world.[2] This is why QBism prefers to talk about an (active) agent rather than a (passive) subject/observer. For CQR, the quantum agent/subject/observer uses QM as a W-rule (norm) in reality, both practically and theoretically. (Let us remind that a W-rule is a rule rooted in its language games, experience, and reality.) One could also say that this is the use of reality with the help of an ideal W-rule (norm). The "friction" between the ideal and the real allows us to speak of the existence (objectivity) of the ideal (but not its real existence; with some stretch, one could only say that the ideal is real in a "weak sense" – in the sense of its rootedness in reality), as well as the use of the real: the use of the ideal, which is rooted in the real, is the use of the real, which is simply a use (a language game) in which the ideal defines, delineates, and identifies the real, and thereby, as it were, becomes identical with it.

Suppose we perform a measurement on a quantum system whose state is described/represented by a given wave function. The obtained result of the quantum measurement allows for the actualization of the quantum system's wave function. From the perspective of CQR, such actualization simply means fixing an individual, singular context (before the measurement, the context of the preliminary experimental preparation was fixed). No collapse as a physical process occurs.

---

[2] Bitbol gives a transcendental-phenomenological interpretation of QBism, assuming that QBists are speaking allegorically. We do not think so, but we recognize that a transcendental-phenomenological tendency in QBism does indeed exist. It was noted, for example, by J. Pienaar [Pienaar, 2021].

On the one hand, the wave function is a W-rule/norm (or a component of QM, including the Born rule, as a W-rule). It (before measurement) encodes the probabilities of possible measurement outcomes, but not as "objective probabilities" of the external world. Instead, it acts as a norm. In this sense, it does not represent the quantum system or probabilities; it tells us how to conduct research correctly to avoid contradictions and problems (which QBism also states). On the other hand, it fixes the actual state of affairs (before or after the measurement, if it's a wave function actualized as a result of the measurement): it is the eigenfunction of a certain quantum physical quantity (operator). It represents the corresponding state of the quantum system (which QBism denies). In this state, the physical quantity has a value that coincides with its corresponding eigenvalue. The wave function actualized in the "process" of measurement represents the actualized state of the quantum system. But one can also say that it represents a (contextual) aspect of the original – before the measurement – quantum system (wave function). In this sense, for example, measurements of a quantum particle's position and momentum complement each other. Bohr's principle of complementarity is a principle of contextuality.

Thus, from the CQR perspective, the wave function plays a dual role: descriptive and normative. The descriptive role itself is dual: the actualized wave function describes the current state of the quantum system, which can also be treated as an aspect of the quantum system on which the measurement was performed.

QBism also denies representationalism and the collapse of the wave function as a physical process. QBism asserts that QM is not a description of reality (nature) *per se*. It is a prescription (a rule) for making decisions. But QBism also entirely rejects the concept of representing a real quantum system or a real quantum property, denying that the wave function can be seen as describing the real state of a quantum system. In contrast, CQR only rejects global metaphysical (decontextualized) representationalism, according to which a theory is a representation of an external objective reality, meaning it describes a predetermined state of a quantum system located in the external world. At the same time, CQR asserts that the theory describes the state of a quantum system locally – in the context of its observation/measurement. What is measured/observed truly exists, is real, and is as it is measured/observed and as it was before the observation/measurement. Observation/measurement does not create or distort real quantum objects and their properties.

The divergence between CQR and QBism is due to the fact that for QBism, both quantum probabilities and the results of measurement/observation are subjective and relate to the agent's subjective experience. They have no ontic meaning. Thus, there are two levels of subjectivity: the level of probability and the level of observation results.

Regarding the first level, QBism is a variant of subjective (personalist) Bayesianism. That is, QBism appeals to the subjective theory of probability, whose foundations were laid by F. Ramsey and B. de Finetti. For de Finetti, probabilities are not in the external world. Within CQR, we've proposed to interpret these probabilities not as subjective, but as evidential (conditional on one's total evidence as one's total knowledge in context) probabilities in the sense of T. Williamson [Williamson, 2002]. In any case, from the CQR perspective, quantum probabilities are not in the external world, meaning they are not objective in a metaphysical sense. But they are not subjective either (in a metaphysical sense). Together with the wave function, they depend on the context. That is, they become objective as a result of their objectification in the context.[3] In this regard, we note that from the point of view of the "quasi-realist" S. Blackburn, whose position we have previously interpreted as CR, de Finetti's probabilities should not be considered subjective [Blackburn, 2013; Pris, 2022, part 9, ch. 2].

So, for QBism, the wave function encodes the subjective probabilities of obtaining certain subjective measurement results in an individual (subjective) experience. Also for QBism, the Hamiltonian, the unitary evolution operator, and other elements of QM formalism are subjective. This implies that the concept of individual experience is primary in QBism. QBism is a "subjective experience first" approach that assumes a dualism of external (material) and internal (subjective) worlds. QBists also talk about a first-person perspective, which can be interpreted phenomenologically. In fact, many authors, including QBists themselves, establish close ties between phenomenology (of E. Husserl, M. Merleau-Ponty, etc.) and QBism. At the same time, QBism positions itself as a type of realism. In contrast, CR criticizes traditional phenomenologies for being anti-realistic approaches. CQR criticizes phenomenological interpretations of QM and, in particular, the phenomenological interpretation of QBism [Pris, 2022; Pris, 2023].

CQR overcomes QBism's dualism by operating with the concepts of reality, rather than the external world (the world is already a part of reality conceptualized with the help of a theory), and language game, rather than subjective experience. For CQR, non-conceptualized sensory experience is primary, and it has no conceptual (intentional, normative) dimension. This is not subjective experience, but an experience between which and reality there is no distance. This realistic approach also allows us to get rid of the problems that arise from a phenomenological interpretation of QBism.

---

[3] A. Khrennikov also introduces the concept of contextual probability [Khrennikov, 2002, p. 8-9]. We introduced a similar concept independently [Pris, 2023, ch. 3].

Nevertheless, the following important similarity exists: for CQR, the QM formalism, including the wave function and the Born rule, is a W-rule (norm). For QBism, the QM formalism also plays the role of a norm. It is considered an extended decision theory with an additional empirical rule – the Born rule. In other words, quantum formalism is a tool that allows for making an optimal choice. The choice is optimal if it satisfies the QM norm. This similarity between CQR and QBism allows for the re-interpretation of QBism in terms of CQR, that is, to free it from (post-)Modern subjectivism, phenomenalism, and anti-realism [Pris, 2022; Pris, 2023].

Just as for CQR there is no (meta-)rule for following the W-rule (the correctness of its application is justified *post factum*), for QBism, a correct (optimal) quantum decision cannot be reliably derived or predicted. For QBism, the correct decision is the one that *post factum* satisfies QM, including the Born rule.

In CQR, the categorical dualism of theory as a W-rule and reality makes it possible to combine the subject (agent) and the measuring device and treat them as ideal – belonging to the W-rule. In turn, QBism asserts that the measuring device is an extension of the agent. That is, the agent in QBism is the agent in a narrow sense, together with the measuring device. Thus, in both CQR and QBism, there is no intermediary between the quantum agent and reality (the world).

Since QBism retains the concept of the external world, it asserts some "objective truths" about it. One such truth is the Born rule. In other words, by objective truths, QBism understands invariant physical and mathematical structures. For CQR, these belong to the category of the ideal, not the real. But they have an objective existence and allow for grasping and identifying elements of reality. In this sense, QM itself can be considered a "fundamental truth" about physical reality.

Thus, we argue that anti-realistic QBism can be re-interpreted realistically. The correspondence between CQR and QBism can be expressed using a dictionary whose terms are interpreted differently. For example, QM as a W-rule in CQR corresponds to the understanding of QM in QBism as playing the role of a norm for decision-making; primary sensory experience in CQR corresponds to subjective experience in QBism; contextual observation results in CQR, which have both an epistemic and ontic meaning, correspond to subjective measurement results in QBism, which have a purely epistemic meaning; contextuality/indeterminacy in CQR corresponds to subjectivity/ indeterminacy in QBism; the observing and acting observer/agent in CQR corresponds to the acting agent in QBism; the measurement of quantum reality as an application of the W-rule in CQR corresponds to the interaction of the agent with the external world; the objective conceptual structure of QM (the W-rule) in CQR corresponds to objective reality in QBism.

To summarize, QBism is an anti-realistic position. Contextual Quantum Realism (CQR) is a realistic alternative. CQR asserts:

1. A quantum state is not a "personal judgment of the agent" (QBism), nor is it subjective (QBism), but is objective (CQR). It describes not a lived experience (QBism), but the state of a physical system in a context (CQR).

2. A quantum measurement is (in the literal sense) a measurement of quantum reality (CQR), not an action of the agent on the external world (QBism). A quantum measurement can be viewed as an action only in the sense that a cognitive Wittgensteinian language game is an action (CQR).

3. The result of a quantum measurement is objective, although sensitive to context (CQR), not subjective (QBism) and relative (QBism), nor is it personal to the agent performing the action (QBism).

4. The quantum formalism is normative (CQR and QBism) and at the same time descriptive and representational locally, in context (not in the sense of representation in an internalist sense) (CQR). The wave function tells what to expect and how to conduct a quantum experiment (it plays the role of a norm), and it also describes, identifies (and in this sense "represents") the state of a quantum system in a context (CQR).

5. Unitary evolution is objective (CQR), not subjective (QBism). It does not express the agent's degree of belief (QBism).

6. A probability of 1 is a judgment of an ontic nature (CQR), not the maximum degree of the agent's subjective confidence without ontic content (QBism).

7. In the general case, the measurement results are not predetermined (CQR and QBism), that is, "unperformed experiments have no results" [Peres, 1978, p. 745] (for CQR this is an analytic judgment, while for QBism it is a substantive thesis), but they are predetermined in the case of probabilities of 1 and 0 (CQR). In the case of probabilities of 0 and 1, one can speak of performed measurements (CQR).

8. Quantum theory is a universal W-rule (norm), that is, a rule (norm) rooted in experience, in reality (CQR), that can be used by any competent subject and is applicable to both the microcosm and the macrocosm (CQR and QBism).

## 4 CQR vs. RQM

### 4.1 Copenhagenish Interpretations

QBism, RQM, and the Copenhagen, pragmatist, and informational interpretations are all classified as "Copenhagenish interpretations", characterized by four properties: 1. "Observers observe" (definite outcomes when making measurements). 2. Universality ("quantum theory is a fundamental physical theory (…) anything in the universe (…) can in principle be described by quantum theory"). 3. "Anti-ψ-ontology" (epistemic character of the wave function). 4. Completeness (there are no hidden variables) [Schmid, Ying, Leifer, 2025].

CQR interprets these characteristics not substantively but analytically – as the "grammar", or the conceptual structure of quantum theory. Furthermore, CQR argues that this structure is inherent to any well-established and empirically confirmed theory that functions as a W-rule [Pris, 2023].

In short, the result of a physical quantity's measurement (the application of the theory as a W-rule) is definite/unambiguous by definition. The wave function has an epistemic character in the sense that the quantum theory that uses it (like any genuine theory) provides knowledge, but not knowledge about a predetermined (external) world. The ontic (objective) is secondary, though the real is primary. For CQR, this is authentic knowledge of the world, of reality, as the correct application of a W-rule implies contact with reality. There are no hidden variables, as a genuine theory cannot fail to be complete. (The W-rule is applied in a context; its application does not require a meta-rule for its own application.) Finally, quantum mechanics, like any theory, is universal within its domain of applicability (the W-rule has a domain of applicability, and its use outside this domain is meaningless) [Pris, 2023]. In a similar vein, Einstein spoke of a "principle theory", and Heisenberg used the term "closed theory".

A key distinguishing feature of both RQM and QBism is that they consider facts to be relative, defined in relation to an "observer". However, they interpret both the observer and, consequently, relativity itself, differently.[4]

---

[4] See our critique of the concept of relative fact in [Pris, 2023, ch. 2].

## 4.2 RQM

Rovelli highlights five "key features" of RQM [Rovelli, 2021a, web, p. 1–2].

1. "(a) The interpretation is realist in the sense that it describes the world as a collection of real systems interacting via discrete relative quantum events." [Rovelli, 2021a, web, p. 1–2].

It is unclear whether this refers to a fundamental ontology of physical systems or events. In the first case, we have a metaphysical realism of the external world, and the concept of a physical system becomes absolute, not relative. In the second, we have a kind of relationism/correlationism, where events are the result of correlations (literally, quantum correlations), i.e., interactions of physical systems where a division into separate independent systems is impossible. The claim of a new realistic approach seems unjustified.

2. "(b) The wave function is interpreted epistemically" [Rovelli, 2021a, web, p. 2].

In RQM, it functions as a mathematical tool with physical meaning for acquiring knowledge. This brings RQM closer to QBism and CQR.

3. "(c) There are no special systems playing the role of 'observers', no special role given to 'agents', or 'subjects of knowledge', no fundamental role given to special 'measurement' contexts" [Rovelli, 2021a, web, p. 2]. "Relative facts: Events, or facts, can happen relative to any physical system" [Adlam, Rovelli, 2022, web, p. 2].

This is a purely physicalist approach that, unlike CQR, refuses to consider the normative, epistemic, and mental dimensions, i.e., the normative practice of applying a theory. According to CQR, these dimensions are ideal. The observer/subject, if not themselves observed, and the observation/measurement belong to the category of the ideal. There is no causal/physical influence of the observer on the observed system.[5] Observation is "reading" information, gaining knowledge. But this reading is contextual, which can be mistakenly taken for a disturbing influence of the observer on the observed physical system.

While "context" is often treated as a technical (operational) concept in quantum mechanics, within CQR, acknowledging it means acknowledging the normative dimension of the theory. The theory as a W-rule (norm) does not describe a predetermined reality but prescribes how to identify its elements. A norm can be applied correctly or incorrectly – always in a context that is implied by its application and is not external to it.

---

[5] Bohr pointed out the logical distinction between the observation instrument and the observed system. According to the Danish physicist, it is wrong to think that quantum observation disturbs the observed phenomena.

4. "(d) The traditional tension between unitary evolution and wave function collapse is resolved by relativising values" [Rovelli, 2021a, web, p. 2].

The similarity between RQM, CQR, and QBism is that problems like the collapse of the wave function, instantaneous action at a distance, drawing a dividing line between the microscopic/quantum and macroscopic/classical, and the Wigner's friend paradox do not arise. In RQM (and QBism), they don't arise because quantum events are "relative facts". In CQR, this is thanks to acknowledging that ontology is sensitive to context [Pris, 2020, ch. 11]. This is not a verbal difference; it implies a different concept of reality and the categorical dualism of the ideal and the real.

5. "(e) A coherent picture of the world is provided by all the values of variables with respect to any single system; juxtaposing values relative to different systems generates apparent incongruences, which are harmless because they refer to a non-existing 'view from outside the world'" [Rovelli, 2021a, web, p. 2]. An RQM postulate states: "Relativity of comparisons: it is meaningless to compare the accounts relative to any two systems except by invoking a third system relative to which the comparison is made" [Adlam, Rovelli, 2022, web, p. 2]. RQM, like CQR, explicitly rejects metaphysical realism. What remains, however, is not realism but relationism/correlationism. Therefore, some authors have interpreted RQM in neo-Kantian terms.

RQM is a purely physicalist (objectivist) approach that refuses to consider epistemic, mental, and truly normative dimensions. For Rovelli and his co-authors, "knowledge" is not epistemic but physical, arising from the interaction and entanglement of quantum systems. RQM either absolutizes physical systems or absolutizes the relations between them, thus failing to maintain its declared relationality. Alternatives to such hidden fundamentalism (stopping at an arbitrary assertion/belief) are coherentism (a circular justification) and infinitism (an infinite regress), which can also be found in works on RQM. A vicious circle arises when the concept of "relation between systems" presupposes that the systems themselves are defined through relations, and an infinite regress occurs when the relation between systems (quantum entanglement) is considered relative to a third system, which then interacts (entangles) with the first two, and therefore requires the introduction of a system relative to which this entanglement is considered, and so on, *ad infinitum*. This is an analogue of the Agrippa's trilemma in the theory of knowledge, which, in our opinion, is a consequence of ignoring the normative dimension and naturalizing the ideal (normative). RQM does not solve the main philosophical problem of quantum mechanics – the measurement problem – and it faces the problem of intersubjectivity of knowledge.

## 5 Final Remarks

Neither RQM nor QBism are truly contextual or fully relational approaches. RQM consciously gets rid of epistemic concepts, ontologizing them, which means it mistakes the ideal (epistemic) for the real (ontic). QBism, on the other hand, appeals to subjectivity instead of real sensory experience, or to the transcendental-phenomenological constitution of objectivity and intersubjectivity.

Rovelli views RQM as a "democratization" of the Copenhagen interpretation. We argue for an additional step: a contextual democratization of RQM. This is achieved by accounting for the categorical distinction between the ideal and the real, as understood by CQR, which treats theory as a W-rule [Pris, 2023, part 9, ch. 3; Pris, 2024]. Both CQR and QBism agree on the normative character of quantum theory, but for CQR, quantum theory as a W-rule also has a local representative dimension, describing and identifying the state of a quantum system within a context.

Bitbol criticizes the naturalization of epistemic concepts in RQM [Bitbol, 2024]. He sympathizes with QBism, interpreting it in a transcendental-phenomenological way, prioritizing epistemology and refusing to recognize the primacy of the concept of reality. While CQR is a full-fledged ontological (though not dogmatic, but critical and contextual) realism, Bitbol's position is only an epistemic realism, asserting the objectivity and knowability of truth.

Bitbol writes: "Between QBism and relational quantum mechanics, there is basically the same difference as between a transcendental theory and a naturalized theory of knowledge; there is the same difference as between a theory that makes the act of knowing the condition of possibility for the concepts of objects, and a theory that assimilates the act of knowing to an interaction between objects... *Relational quantum mechanics is to QBism what the pre-critical relationism of the young Kant is to the critical relationism of the mature Kant*" [Bitbol, 2024, p. 4].

CQR rejects both phenomenological and transcendental methods. It embraces a grammatical method and the primacy of reality (instead of experience, lived experience, phenomenon, or others). The concept of lived experience in the sense of Husserl's transcendental phenomenology, which Bitbol bases his position on, actually distances us from reality, as the French phenomenologist C. Romano has shown. It can and should be reinterpreted within the framework of a realistic phenomenology [Romano, 2025]. CQR replaces this concept with that of primary sensory experience, between which and reality there is no distance.

Bitbol is right to note that RQM separates physics from epistemology by replacing epistemic concepts with ontic ones. At the same time, Bitbol himself separates physics from reality, replacing metaphysical (reality) and ontological concepts with epistemic ones and reducing the former to the latter.

The three interpretations and Bitbol's position can be compared based on how they handle the problem of the so-called Heisenberg's cut (in German: *Schnitt*) between the quantum and the classical.

For RQM, the problem doesn't arise because this interpretation is "democratic", treating all physical systems, including the observer, on an equal footing.

From the perspective of QBism in its subjectivist interpretation, the problem also doesn't arise, but for the opposite reason – everything is treated subjectively.

From a transcendental-phenomenological perspective, as Bitbol writes, the cut exists but is a transcendental limit.

From the CQR perspective, the cut is treated in terms of the categorical dualism between the real (sensory, experiential) and the ideal (theoretical). It is defined contextually, or as Heisenberg said, by the nature of the problem. In other words, the Heisenberg cut is a specific manifestation of the general philosophical problem of defining the movable boundary between the real and the conceptual/ideal, which CR points to [Pris, 2023; Benoist, 2024].

Albert Einstein believed that QM was incomplete. And it is indeed incomplete, but not in a mathematical or physical sense. It is philosophically incomplete if its interpretation – which is inseparable from the theory itself – remains within the framework of (post-)Modern philosophy. This is the root cause of the quantum paradoxes. In this sense, CQR makes QM complete, thereby ridding it of its paradoxes.